\documentclass[doublecol,linenumbers]{epl2}

\usepackage{epsfig}
\usepackage{dcolumn}
\usepackage{amsmath}
\begin{document}


\title{\bf Evidence of radiation-driven Landau states in 2D electron systems:
magnetoresistance oscillations phase shift}

\author{Jes\'us I\~narrea$^{1,2}$ }

\institute {$^1$Escuela Polit\'ecnica
Superior,Universidad Carlos III,Leganes,Madrid,Spain and
$^2$Unidad Asociada al Instituto de Ciencia de Materiales, CSIC,
Cantoblanco,Madrid,28049,Spain.}

\pacs{nn.mm.xx}{First pacs description}
\pacs{nn.mm.xx}{Second pacs description}
\pacs{nn.mm.xx}{Third pacs description}

\abstract
{We provide the ultimate explanation  of one of the core features of microwave-induced magnetoresistance oscillations in high
mobility two dimensional electron systems: the 1/4-cycle
phase shift of minima.   We start with the radiation-driven electron orbits model
with the novel concept of scattering flight-time between Landau states.
We calculate the extrema and nodes positions obtaining an exact coincidence with
the experimental ones. The main finding is that the physical origin
of the phase shift is a delay of $\frac{\pi}{2}$ of the  radiation-driven Landau guiding center with
respect to radiation, demonstrating the oscillating nature of the
irradiated Landau states.
 We analyze the dependence of this
minima  on radiation frequency and  power and its
possible shift with the quality of the sample.}

\maketitle
\section{Introduction}
Microwave-induced magnetoresistance ($R_{xx}$) oscillations (MIRO)\cite{mani1,zudov1}, show up in  high mobility
two-dimensional electron systems (2DES) when they are irradiated with microwaves (MW) at low
temperature ($T\sim 1K$) and under low magnetic fields ($B$) perpendicular to the 2DES.
At high enough MW  power ($P$) maxima and
minima oscillations increase but the latter evolve into zero resistance states (ZRS)\cite{mani1,zudov1}.
Both effects  were totally unexpected when they were first obtained revealing some
type of new radiation-matter interaction or coupling assisting  electron magnetotransport\cite{ina1,ina11,ina12}.
Despite that over the last few years important experimental\cite{mani01,mani02,mani2,
mani3,willett,mani4,smet,yuan,mani03,mani04,mani05,mani5,wiedmann1,wiedmann2,kons1,vk,mani6,mani61,mani62,mani7}
and theoretical efforts \cite{ina2,ina21,ina22,ina23,ina24,girvin,lei,ryzhii,rivera,vavilov,ina4,ina41,ina42,ina5,ina51,ina71,ina72}
have been made on MIRO
and ZRS, their physical origin still  remains controversial and far from reaching a definite
consensus among the people devoted to this field. For instance, the two, in principle, accepted
theoretical models explaining MIRO, (displacement\cite{girvin} and inelastic\cite{vavilov} models)
are under question in regards of  recent (and even older) experimental results\cite{mani71,zudov10} that they are
not able to explain. In fact, experimentalists on MIRO are calling for other theoretical approaches that
might be more successful offering solid arguments on  MIRO and ZRS physical origin\cite{lei,ina2,ina21}.



Among the different  features defining MIRO we can underline  some that can be consider as
fundamental, turning up in most experiments irrespective of the semiconductor
platform and carrier (holes or electrons). Thus, we can highlight three of them:
MIRO are periodic in $B^{-1}$, MIRO dependence with $P$ follows a sublinear law whose
exponent is around $0.5$\cite{mani6,mani62} and finally they present a $1/4$-cycle phase shift in the minima
position\cite{mani2,mani4,zudov11}.
 To date, none of the existing theoretical models on MIRO have been able
to  provide a convincing explanation about the minima shift  and
why they depends in such a way on MW frequency ($w$) and $B$ or
cyclotron frequency ($w_{c}$). On the other
hand they are, according to experiments, immune to MW power.


\begin{figure}
\centering \epsfxsize=3.5in \epsfysize=4.0in
\epsffile{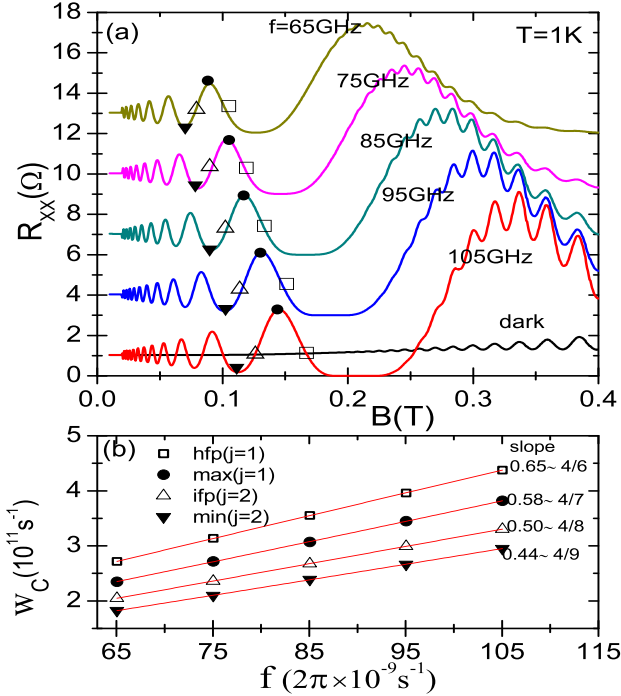}
\caption{Dependence of MIRO extrema and nodes with radiation frequency.
1a) Calculated $R_{xx}$ versus $B$ under radiation for  several frequencies (from 65 GHz to
105 GHz). The dark curve is also presented. Extrema and nodes
displace to higher  $B$ as $w$ increases.  1b) Cyclotron frequency  $w_{c}$ vs radiation frequency
$w=2\pi f$, where $f$
is the frequency in $GHz$, for the extrema and nodes labelled in the upper panel.
For the hifp, the square symbol, for the maxima, the black dot   symbol,
for the ipf, the triangle  symbol
and finally, for the minima, black down triangle. In panel we plot
also the fits of the four sets of data giving straight lines. The slopes
correspond with the extrema and nodes positions of the upper panel.
}
\end{figure}

In this letter, we present,  based on the
radiation-driven electron orbits model\cite{ina2,ina21,ina22}, a theoretical analysis on MIRO where
we explain the 1/4-cycle shift of minima and the peculiar position of the
extrema and nodes.
According to this
model, when a Hall bar is illuminated, the
guiding centers of the Landau states perform a classical trajectory consisting in a harmonic
motion along the direction of the current. Thus, the electron orbits move in phase and
harmonically with each other  at the radiation frequency, altering dramatically the scattering
conditions and giving rise eventually to MIRO and, at higher $P$, ZRS.
Now, our approach consists on two
main physical effects. The first is that the Landau orbit guiding center
displacement lags behind the driving force (radiation) by a definite phase constant of $\frac{\pi}{2}$.
And secondly, in a scenario of remote
charged impurity scattering, we introduced the concept of scattering
{\it flight time} as the time it takes the electron to jump from one Landau orbit to another
in a scattering process. The core finding is that this time is equal to
the cyclotron period. Thus, during the flight time the electrons
complete one loop in their cyclotron orbits. For the second effect to happen
it is essential that magnetotransport
is dominated by the guiding center position of the Landau orbit.

In our calculations we recover the experimental expressions for the extrema and nodes
positions showing that only depend on $w$ and $w_{c}$.
We finally discuss and predict the situation of a possible change in the minima shift
for ultrahigh mobility samples when ramping up the magnetic field. The reason is that
the phase constant delay of the MW-driven guiding center with respect to radiation,
 depends on
the  damping that the Landau orbits suffer along the swinging motion. This damping
drops in samples with extreme high mobility (much lower disorder), affecting the phase constant (delay) and
eventually the extrema positions and minima shift.

\section{ Theoretical Model}
The {\it radiation-driven electron orbits model}, was proposed   to study
the magnetoresistance of a 2DES subjected to MW at low $B$ and temperature, $T$\cite{ina2,ina21,ina30,kerner,park}.
The total electronic hamiltonian $H$ can be
exactly solved
and the solution for the total
wave function of $H$\cite{ina2,ina30,kerner,park,ina4}reads:
$\Psi_{n}(x,t)\propto\phi_{n}(x-X_{0}-x_{cl}(t),t)$,
where $\phi_{n}$ is the solution for the
Schr\"{o}dinger equation of the unforced quantum harmonic
oscillator. Thus, the  obtained wave
function (Landau state or Landau orbit) is the same as the one of the standard quantum harmonic oscillator where the guiding
center of the Landau state, $X_{0}$ without radiation, is displaced by $x_{cl}(t)$.
$x_{cl}(t)$ is the classical solution of a negatively charged, forced  and damped, harmonic
oscillator\cite{french,main}:
\begin{eqnarray}
x_{cl}(t)&=&\frac{-e E_{o}}{m^{*}\sqrt{(w_{c}^{2}-w^{2})^{2}+\gamma^{4}}}\cos ( wt-\beta)\nonumber\\
&=&-A\cos ( wt-\beta)
\end{eqnarray}
 where $E_{0}$ is the amplitude of the MW electric field and $\beta$ is a phase constant.
$\beta$ is the phase difference between the radiation-driven guiding center and
the driving radiation; its expression reads, $\tan \beta= \frac{\gamma^{2}}{w_{c}^{2}-w^{2}}$.
Thus, the guiding center lags behind the time-dependent driving force, (radiation), a phase
constant of $\beta$. In the above calculations radiation has been expressed as $E(t)=E_{0} \cos wt$.
 $\gamma$ is a phenomenologically-introduced damping factor
for the  interaction of electrons with the lattice ions
giving rise to the emission of acoustic phonons.
When the damping parameter $\gamma$ is important, ($\gamma > w\Rightarrow \gamma^{2}>> w^{2}$), then $\tan \beta\rightarrow \infty$
 and $\beta\rightarrow \frac{\pi}{2}$.
Now, the time-dependent guiding center is, $X(t)=X_{0}+x_{cl}(t)=X_{0}-A \sin wt$.
This physically implies that the   orbit guiding centers oscillate harmonically at the MW frequency,
but radiation leads the guiding center displacement in $\frac{\pi}{2}$.
This expression automatically fulfills the initial condition of
$X(t=0)=X_{0}$ and then we do not need to add any initial phase.

This $radiation-driven$ behavior of the orbit guiding centers is supposed to  affect  the
electron-charged impurity scattering and eventually  $R_{xx}$\cite{kerner,miura,ina6}.
If one electron,  without radiation, is scattered from the Landau state $\Psi_{n}$ located at $X_{0,n}$  to the final state
$\Psi_{m}$ at  $X_{0,m}$ in a time $\tau$, the average advanced distance, is given by $\Delta X_{0} = X_{0,m}-X_{0,n}$.
 This  {\it flight time}, $\tau$, is  the time it
takes the electron to "fly"  from one orbit to another due to scattering. This
time is part of the quantum scattering time, (quantum lifetime), $\tau_{q}$,
 that it is normally defined as
the average time between scattering events or collisions, considering that all of them
are equally weighted.
 In high mobility 2DES the electron magnetotransport
is ruled or dominated  by the guiding center position of the Landau orbit
and not by the electron position itself. In a semiclassical approach, this
 implies that the electron scattering begins and ends
in the same relative position inside the initial and final cyclotron orbits.
In other words, during the scattering jump from one orbit to another, in a time $\tau$,
the electrons in their orbits complete one full loop. Then, $\tau$ must be equal to $T_{c}$:
$  \tau = \frac{2\pi}{w_{c}}=T_{c}$ where $T_{c}$ is the cyclotron period.
Interestingly, applying the
time-energy uncertainty relation\cite{cohen} $\Delta t\cdot \Delta E \geq h$,
it turns out that, being $\Delta t =\tau$,
then, $\tau \times \Delta E = \frac{2\pi}{w_{c}} \times \Delta E \geq h\Rightarrow$
  the uncertainty of  energy is, $\Delta E  \simeq \hbar w_{c}$. Then the scattered electron
ends up, most likely, in the next Landau level: $m=n+1$.

When we turn on  radiation, the guiding centers begin to harmonically oscillate
according to the expression of $X(t)$. Later, in a time $t_{i}$,  scattering
starts and the electron jumps from  $\Psi_{n}(t)$ at $X_{n}(t)$ to $\Psi_{m}(t)$ at  $X_{m}(t)$ in a
total time $t_{f}=t_{i}+\tau$. The average advanced distance will change accordingly:
 $\Delta X(t)=X_{m}(t_{f})-X_{n}(t_{i})=\Delta X_{0}-A \sin w(t_{i}+\tau)+A \sin wt_{i}$.
The time $t_{i}$ is on average the time it takes the electron to
experience a  scattering proces, i.e., the scattering time $\tau_{q}$, that for constant $B$ it is
constant too. Therefore,  $t_{i}\simeq \tau_{q}$ and $\Delta X(t)=\Delta X_{0}-A \sin (\tau +\phi)+A \sin \phi$,
being $\phi= w\tau_{q}$. Now shifting the time origin so that $\phi=0$, i.e., to when
the electron scattering begins, we obtain the final expression:
 $\Delta X(t)=\Delta X_{0}-A \sin w\tau$.\\
The longitudinal conductivity $\sigma_{xx}$
is given by\cite{ridley}
$\sigma_{xx}\propto \int dE \frac{(\Delta X^{MW})^{2}}{\tau_{q}}$
being $E$ the energy
 and $\frac{1}{\tau_{q}}$ is  the
remote charged impurity scattering rate\cite{ina2}.
To obtain $R_{xx}$ we use the common tensorial relation
$R_{xx}=\frac{\sigma_{xx}}{\sigma_{xx}^{2}+\sigma_{xy}^{2}}
\simeq\frac{\sigma_{xx}}{\sigma_{xy}^{2}}$, where
$\sigma_{xy}\simeq\frac{n_{i} e}{B}$, $n_{i}$ being the electrons density, and $\sigma_{xx}\ll\sigma_{xy}$.
Thus\cite{ina6},
$R_{xx}\propto - A  \sin w\tau$.
\begin{figure}
\centering\epsfxsize=3.5in \epsfysize=4.0in
\epsffile{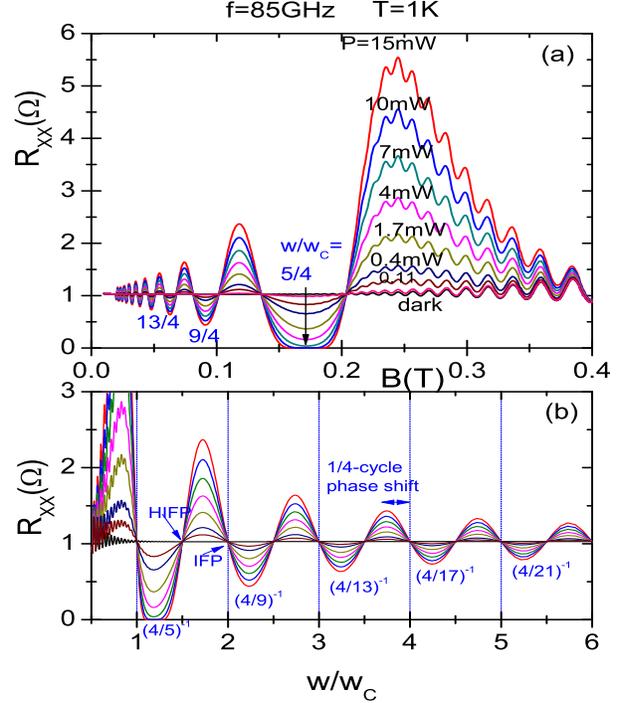}
\caption{Dependence of the extrema and nodes positions
of MIRO with the radiation power. 2a) Calculated magnetoresistance, $R_{xx}$, irradiated
with MW of $85$  GHz vs $B$ for different values of $P$, from $15$ mW to dark
and $T=1$ K. We observe how magnetoresistance oscillations tend to vanish
as the power decreases.  2b) Same results as in 2a but this time vs
$w/w_{c}$. In both panels we mark the values defining the 1/4-cycle phase
shift. The main result is that the positions of extrema and
nodes are immune with respect to the power.
}
\end{figure}

Importantly, according to the last expression, the MIRO minima positions are given by:
$w\tau = \frac{\pi}{2}+2\pi j \Rightarrow w = \frac{2 \pi}{\tau}\left(\frac{1}{4}+ j \right)$,
$j$ being a positive integer.
And for the MIRO maxima:
$w\tau = \frac{3 \pi}{2}+2\pi j \Rightarrow w = \frac{2 \pi}{\tau}\left(\frac{3}{4}+ j \right)$.
We can obtain also expressions for MIRO nodes, or the points
where the radiation curve crosses the dark curve.
The right nodes, also known as {\it half integer fixed points} (hifp), fulfill, $w = \frac{2 \pi}{\tau}( j+1/2)$. And the left ones, known as
{\it integer fixed points} (ifp), $w = \frac{2 \pi}{\tau} j$.
If we compare the calculated expression  with the ones
 previously obtained in experiments by Mani et al\cite{mani2,mani4}:
\begin{eqnarray}
theory    &  \Longleftarrow \Longrightarrow  &experiment \nonumber\\
min\rightarrow w = \frac{2 \pi}{\tau}\left(\frac{1}{4}+ j \right)&\Leftrightarrow& \frac{w}{w_{c}} = \left(\frac{1}{4} + j\right)\nonumber\\
max\rightarrow w = \frac{2 \pi}{\tau}\left(\frac{3}{4}+ j \right)&\Leftrightarrow& \frac{w}{w_{c}} = \left(\frac{3}{4} + j\right)\nonumber\\
hifp\rightarrow w = \frac{2 \pi}{\tau}\left(j+\frac{1}{2} \right)&\Leftrightarrow& \frac{w}{w_{c}} = \left(j+\frac{1}{2} \right)\nonumber\\
ifp\rightarrow w = \frac{2 \pi}{\tau} j &\Leftrightarrow& \frac{w}{w_{c}} =j  \nonumber\\
 \nonumber\\
\end{eqnarray}
 and thus,
$  \tau = \frac{2\pi}{w_{c}}=T_{c}$.
This interesting result
 confirms our previous approach about the physical insight of the
 flight time.
Now, the 1/4-cycle phase shift of the MIRO minima can be traced back to a
phase constant of $\frac{\pi}{2}$ that is the delay that presents the driven guiding center
with respect to radiation in the whole range of $B$. This delay is revealed by the
interplay between the swinging nature of the Landau states under radiation and the remote
charged impurity scattering. Then, we can state that  the 1/4-cycle
phase shift in MIRO minima is an evidence of that the Landau states are not fixed but
they oscillate driven by MW.
The final expression of the
irradiated magnetoresistance  turns out to be:
\begin{equation}
R_{xx}\propto -  \frac{e E_{o}}{m^{*}\sqrt{(w_{c}^{2}-w^{2})^{2}+\gamma^{4}}} \sin \left(2\pi\frac{w}{w_{c}}\right)
\end{equation}

\begin{figure}
\centering\epsfxsize=3.5in \epsfysize=3.5in
\epsffile{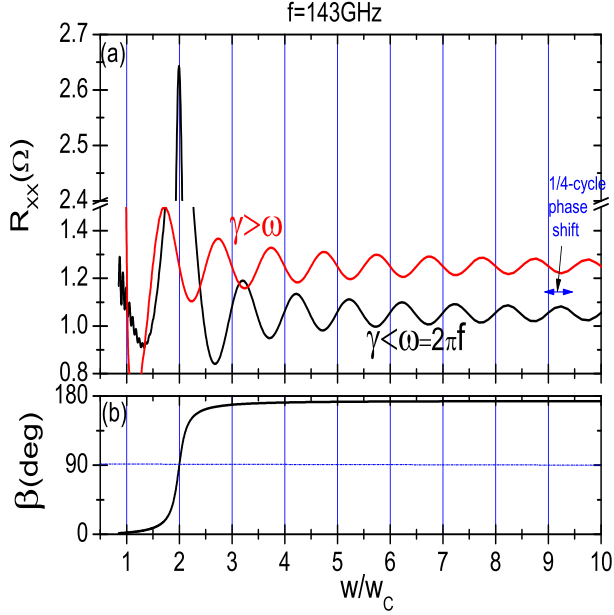}
\caption{Calculated magnetoresistance calculated results of irradiated $R_{xx}$ and $\beta$ vs $w/w_{c}$
for a regime of $\gamma< w$ and a frequency of $143$ GHz. As in Y. Dai experiments, this simulations has been run
considering that the resonance takes place in the second harmonic, i.e., $2w_{c}=w$.
This permits in the simulation to see more clearly the transition of $\beta$ from $\pi$ to $\frac{\pi}{2}$.
}
\end{figure}

\section{Results.}
In Fig.1 we exhibit the dependence of MIRO extrema and nodes with respect to $w$.
In Fig.1a, we show calculated irradiated $R_{xx}$ versus $B$ for  an intermediate range of $w$.
We observe, as expected according to the set of equations (3), that extrema and nodes
displace to higher  $B$ as $w$ increases. In Fig. 1b, we plot $w_{c}$ vs $w=2\pi f$, where $f$
is the radiation frequency in $GHz$, for the extrema and nodes labelled in Fig. 1a.
For the hifp we have used the square symbol, the index $j=1$ and then from equation (3) we
obtain, $w_{c}=w\times\frac{4}{6}$.
For the maxima, the black dot   symbol, the index $j=1$ and then  $w_{c}=w\times\frac{4}{7}$.
For the ipf, the triangle  symbol, $j=2$ and $w_{c}=w\times\frac{4}{8}$.
And finally, for the minima, black down triangle, $j=2$ and $w_{c}=w\times\frac{4}{9}$. These calculated
results demonstrate and explain that the corresponding shifts for extrema and nodes are
independent of radiation frequency. They only depend on the phase difference
between radiation and the harmonic displacement of the Landau orbit
guiding center. In the lower panel we present the fits corresponding to
the four set of data resulting, as expected, straight lines where the slopes
are according to the extrema and nodes positions in the upper panel.

In Fig. 2, we present the $P$ dependence of the extrema and nodes positions
of MIRO. In Fig. 2a we plot $R_{xx}$ irradiated
with MW of $85$  GHz vs $B$ for different values of $P$, from $15$ mW to dark
and $T=1$ K. We observe how MIRO decrease and tend to vanish
as $P$ drops. In Fig. 2b, we exhibit same results as in 2a but this time vs
$w/w_{c}$. In both panels we mark the values defining the 1/4-cycle phase
shift. The main finding is that the positions of extrema and
nodes turn out to be immune with respect to $P$.
Now, we can theoretically explain these results according to our model.
In equation (3) $P$ only shows up in the numerator of the amplitude as $\sqrt{P}\propto E_{0}$,
and not in the phase of the sine function.
Thus, $P$ does not affect the phase of MIRO  and the latter  keeps constant.
The outcome is that extrema and node positions in MIRO turn out to be immune to $P$.

Finally it is interesting to consider the case when the damping parameter
$\gamma$ is small compared to $w$. Under this condition we will come across with regimes
where the 1/4-cycle phase shift will not be conserved. This can be found in ultra-high
mobility samples  where the damping parameter is expected to get smaller due to a
much smaller disorder.
 Thus, $\gamma < w\Rightarrow \gamma^{2}<< w^{2}$, then the obtained values
 for $\beta$ will be different from before when ramping up $B$.
Thus, if along with this condition, $w_{c}\simeq 0$, (low values of $B$), then  $\tan \beta\rightarrow 0$
 and $\beta\rightarrow \pi  \Rightarrow X(t)=X_{0}-A \cos( wt -\pi)$. And
 we need to add an initial phase constant of $-\frac{\pi}{2}$ to fulfill
 $X(t=0)=X_{0}$. Finally $X(t)=X_{0}-A \cos( wt -\frac{3\pi}{2})=X_{0}+A \sin wt $.
 Similarly as before, this result leads to $\Delta X(t)=\Delta X_{0}+A \sin w\tau$.
 According to this  we would expect different positions for extrema and nodes. For instance, now for maxima:
 $w\tau = \frac{\pi}{2}+2\pi j \Rightarrow w = \frac{2 \pi}{\tau}\left(\frac{1}{4}+ j \right)$.
 And for minima: $w\tau = \frac{3 \pi}{2}+2\pi j \Rightarrow w = \frac{2 \pi}{\tau}\left(\frac{3}{4}+ j \right)$.
 These are opposite results to the ones obtained in the previous scenario of $\gamma>w$.
 If now $B$ and $w_{c}$ are  increased, $\tan \beta$ increases too,
  but in negative and tends to $-\infty$ at resonance,
where $\beta\simeq \frac{\pi}{2}$. Accordingly, $ X(t)=X_{0}-A \cos( wt -\frac{\pi}{2}-\frac{\pi}{2})=X_{0}+A \cos wt$,
and now  maxima positions are given by,  $w\tau = 2\pi j$.
If $w_{c}$ keeps increasing, $\tan \beta\rightarrow 0$
but now $\beta\rightarrow 0$. Thus, as $B$ is ramping up from $B=0$,  $\beta$ evolves from $\pi$ to
$\frac{\pi}{2}$ at resonance and, for higher  $B$, ends up being equal to $0$.
Something similar to the  above has been experimentally obtained previously
 by Y. Dai et al.\cite{dai} but it has been overlooked by the people, experimentalist and
theorist, devoted to MIRO; they obtain a clear shift of extrema and nodes positions as
$B$ rises for ultra-high mobility samples.

In Fig. 3 we present calculated results of irradiated $R_{xx}$ and $\beta$ vs $w/w_{c}$
for a regime of $\gamma< w$ and a frequency of $143$ GHz. As in Y. Dai experiments, this simulations has been run
considering that the resonance takes place in the second harmonic, i.e., $2w_{c}=w$.
This permits in the simulation to see more clearly the transition of $\beta$ from $\pi$ to $\frac{\pi}{2}$.
In the upper panel we plot two curves, one for $\gamma< w$ (black curve) an the
other for $\gamma> w$ (red curve) to compare both regimes in terms of
extrema and nodes positions as $B$ rises. We observe that the black curve does
not longer shows the 1/4-cycle phase shift for minima. Yet it presents
this shift but for the maxima and changes to $w=w_{c} j$ when approaching to
$w=2w_{c}$. In the lower panel it can be observed that the transitions persists
up to $\beta\simeq 0$ as $B$ keeps increasing.

\section{Conclusions}
In summary, we have presented a theoretical analysis
of one of the main features of MIRO as is the 1/4-cycle phase shift
of minima. We have used the radiation driven electron orbits model
introducing the concept of flight time between Landau state
in a scattering process. The first finding is that this time is
equal to the cyclotron period. We have obtained an exact coincidence between
the calculated values for extrema and node positions and the experimental ones.
The key outcome of our analysis is that this shift is an evidence
of a delay of $\frac{\pi}{2}$ between the oscillating Landau state guiding center and radiation.
In consequence,  the Landau states oscillates harmonically  when irradiated and it is the
interplay of this effect with scattering by charged remote impurities that gives
rise of the appearance of MIRO.
We have studied also the conditions so that this minima shift can be altered
in ultra-high mobility samples.

\section{Acknowledgments}

This work is supported by the MINECO (Spain) under grant
MAT2014-58241-P  and ITN Grant 234970 (EU).
GRUPO DE MATEMATICAS APLICADAS A LA MATERIA CONDENSADA, (UC3M),
Unidad Asociada al CSIC.



\begin{thebibliography}{19}

\bibitem{mani1} R. G. Mani, J. H. Smet, K. von Klitzing, V. Narayanamurti,
W. B. Johnson, and V. Umansky, Nature(London) \textbf{420}, 646
(2002)

\bibitem{zudov1} M. A. Zudov, R. R. Du, L. N. Pfeiffer, and K. W. West,
Phys.Rev. Lett. \textbf{90}, 046807 (2003)


\bibitem{ina1}
J. I\~narrea, G. Platero, Phys. Rew. B, {\bf 51}, 5244, (1995)

\bibitem{ina11}
J. I\~narrea, R. Aguado, G. Platero, Europhys
Lett.  {\bf 40}, 417, (1997)

\bibitem{ina12}
J. I\~narrea, G. Platero, Europhys. Lett.,
{\bf 34}, 43, (1996)
\revision{
\bibitem{mani01}
A. N. Ramanayaka, R. G. Mani, J. Inarrea, and W. Wegscheider,
Phys. Rev. B, {\bf 85}, 205315, (2012)

\bibitem{mani02}
R. G. Mani, V. Narayanamurti, K. von Klitzing, J. H. Smet, W. B. Johnson, and V. Umansky,
Phys. Rev. B, {\bf 69}, 161306(R), (2004)}



\bibitem{mani2}R. G. Mani, J. H. Smet, K. von Klitzing, V. Narayanamurti, W. B. Johnson, and V. Umansky
, Phys. Rev. Lett. \textbf{92}, 146801
(2004).

\bibitem{mani3} R. G. Mani, J. H. Smet, K. von Klitzing, V. Narayanamurti, W. B. Johnson, and V. Umansky
, Phys. Rev. B\textbf{69}, 193304
(2004).

\bibitem{willett} R. L. Willett, L. N. Pfeiffer, and K. W. West, Phys.
Rev. Lett. \textbf{93}, 026604 (2004).

\bibitem{mani4}R. G. Mani, Physica E (Amsterdam) \textbf{22}, 1 (2004);

\bibitem{smet} J. H. Smet, B. Gorshunov, C. Jiang, L. Pfeiffer, K. West,
V. Umansky, M. Dressel, R. Meisels, F. Kuchar, and K. von Klitzing
, Phys. Rev. Lett. 95, 118604 (2005).

\bibitem{yuan} Z. Q. Yuan,C.L. Yang, R.R. Du, L.N.Pfeiffer and
K.W. West , Phys. Rev. B\textbf{74}, 075313 (2006).

\revision{
\bibitem{mani03}
Tianyu Ye, Han-Chun Liu, Zhuo Wang, W. Wegscheider and Ramesh G. Mani,
Sci. Rep. \textbf{5}, 14880, (2015)

\bibitem{mani04}
R. G. Mani, and A. Kriisa,
Sci. Rep. \textbf{3}, 3478, (2013)

\bibitem{mani05}
R. G. Mani,
Appl. Phys. Lett. \textbf{85}, 4962, (2004)}






\bibitem{mani5} R. G. Mani, W.B. Johnson, V.Umansky, V. Narayanamurti and
K. Ploog, Phys. Rev. B\textbf{79}, 205320
(2009).

\bibitem{wiedmann1} S. Wiedmann, G.M. Gusev, O.E. Raichev, A.K. Bakarov, and
J.C. Portal, Phys. Rev. Lett., {\bf 105}, 026804, (2010)

\bibitem{wiedmann2} S. Wiedmann, G.M. Gusev, O.E. Raichev, A.K. Bakarov, and
J.C. Portal, Phys. Rev. B, {\bf 81}, 085311, (2010)

\bibitem{kons1}
D. Konstantinov and K. Kono, Phys. Rev. Lett. {\bf 103}, 266808
(2009)



\bibitem{vk}
S. I. Dorozhkin, L. Pfeiffer, K. West K, K. von Klitzing, J.H. Smet JH,
NATURE PHYSICS, {\bf 7}, 336-341, (2011)


\bibitem{mani6} R. G. Mani, C. Gerl, S.Schmult, W. Wegscheider and
V. Umansky, Phys. Rev. B\textbf{81}, 125320, (2010)

\bibitem{mani61}R.G. Mani, A.N.
Ramanayaka and W. Wegscheider, Phys. Rev. B., {\bf 84}, 085308, (2011)

\bibitem{mani62}
Jesus Inarrea, R.G. Mani and W. Wegscheider, Phys. Rev. ,{\bf 82} 205321 (2010)


\bibitem{mani7} R. G. Mani, Int. J. Mod. Phys. B, {\bf 18}, 3473,
(2004); Physica E, \textbf{25}, 189 (2004)



\bibitem{ina2}
J. I\~narrea and G. Platero, Phys. Rev. Lett. {\bf 94} 016806,
(2005)
\bibitem{ina21} J. I\~narrea and G. Platero, Phys. Rev. B {\bf 72} 193414
(2005)
\bibitem{ina22}J. I\~narrea and G. Platero, Appl. Phys. Lett.,  {\bf 89},
052109, (2006)
\bibitem{ina23}J. I\~narrea and G. Platero, Phys. Rev. B,  {\bf 76},
073311, (2007)
\bibitem{ina24}  J. I\~narrea, Appl. Phys. Lett. {\bf 90}, 172118,
(2007)

\bibitem{girvin}
A.C. Durst, S. Sachdev, N. Read, S.M. Girvin, Phys. Rev. Lett.{\bf
91} 086803 (2003)

(2005)

\bibitem{lei}
X.L. Lei, S.Y. Liu, Phys. Rev. Lett.{\bf 91}, 226805 (2003)

\bibitem{ryzhii}
Ryzhii et al, Sov. Phys. Semicond. 20, 1299, (1986)

\bibitem{rivera}
P.H. Rivera and P.A. Schulz, Phys. Rev. B {\bf 70} 075314 (2004)



\bibitem{vavilov}
M.G. Vavilov et. al., Phys. Rev. B, {\bf 70}, 161306(2004)

\bibitem{ina4}
J. I\~narrea and G. Platero, Appl. Phys Lett. {\bf 93}, 062104,
(2008)

\bibitem{ina41}
 J. I\~narrea and G. Platero, Phys. Rev. B, {\bf 78},
193310,(2008)
\bibitem{ina42}
J. I\~narrea, Appl. Phys Lett. {\bf 92},
192113,(2008)


\bibitem{ina5}
Jesus Inarrea and Gloria Platero, Appl. Phys. Lett.
{\bf 95}, 162106, (2009)

\bibitem{ina51}
J. I\~narrea, Appl. Phys Lett. {\bf 90},
262101,(2007)


\bibitem{ina71}
J. Inarrea, G. Platero and A. H. MacDonald,
Physica Stat. Solid. A, {\bf 203},1148, (2006)

\bibitem{ina72}
J. I\~narrea, Appl. Phys Lett. {\bf 100},
242103,(2012)





\bibitem{mani71}
Tianyu Ye, Han-Chun Liu, W. Wegscheider and R.G. Mani, Phys. Rev. B, {\bf 89},
155307,(2014)

\bibitem{zudov10}
 Q. Shi, P.D. Martin, A.T. Hatke, M.A. Zudov, J.D. Watson, G.C. Gardner,
 M.J. Manfra, L.N. Pfeiffer and K.W: West, Phys. Rev. B, {\bf 92},
081405(R),(2015)

\bibitem{zudov11}
MA. Zudov, Phys. Rev. B, {\bf 69}, 041304(R),(2004)


\bibitem{ina30} J. Inarrea and G. Platero, Appl. Physl Lett. {\bf 89},
172114, (2006)

\bibitem{kerner}
E.H. Kerner, Can. J. Phys. {\bf 36}, 371 (1958).

\bibitem{park}
K. Park, Phys. Rev. B {\bf 69} 201301(R) (2004).






\bibitem{french}
A.P. French,  {\it Vibrations and waves.},  W.W. Norton and Company,
New York, 1971.

\bibitem{main}
I.G. Main, {\it Vibrations and waves in Physics.},
Cambridge Unversity Press, 1993.




















\bibitem{miura}
Noboru Miura,  {\it Physics of Semiconductors in High Magnetic Fields.},
Oxford University Press, (2008).

\bibitem{ina6}
Jesus Inarrea and Gloria Platero, Journal of physics. Condensed matter \textbf{27}  415801 (2015)


\bibitem{cohen}
Claude Cohen-Tannoudji, Bernard Diu and Franck Laloe,
{\it Qauntum Mechanics}, John Wiley and sons, New York, (1977).



\bibitem{ridley}
B.K. Ridley. {\it Quantum Processes in Semiconductors}, 4th ed. Oxford
University Press, (1993).










\bibitem{dai}
Yanhua Dai, R.R. Du, L.N. Pfeiffer and K.W. West, Phys. Rev. Lett. {\bf 105} 246802 (2010)

\end{thebibliography}
\end{document}